**Is Downloading this App Consistent with my Values? Conceptualizing a Value-Centered Privacy Assistant**

Sarah E. Carter[1,2,3]

s.carter6@nuigalway.ie

[1] Data Science Institute, National University of Ireland Galway, Galway, Ireland
[2] Discipline of Philosophy, School of History and Philosophy, College of Arts, Social Sciences, and Celtic Studies, National University of Ireland Galway, Galway, Ireland
[3] Science Foundation Ireland Center for Research Training in Digitally-Enhanced Reality (d-real)

**Abstract.** Digital privacy notices aim to provide users with information to make informed decisions. They are, however, fraught with difficulties. Instead, I propose that data privacy decisions can be understood as an expression of user values. To optimize this value expression, I further propose the creation of a value-centered privacy assistant (VcPA). Here, I preliminary explore how a VcPA could enhance user value expression by utilizing three user scenarios in the context of considering whether or not to download an environmental application, the OpenLitterMap app. These scenarios are conceptually constructed from established privacy user groups - the privacy fundamentalists; the privacy pragmatists; and the privacy unconcerned. I conclude that the VcPA best facilitates user value expression of the privacy fundamentalists. In contrast, the value expression of the privacy pragmatists and the privacy unconcerned could be enhanced or hindered depending on the context and their internal states. Possible implications for optimal VcPA design are also discussed. Following this initial conceptual exploration of VcPAs, further empirical research will be required to demonstrate the effectiveness of the VcPA system in real-world settings.

**Introduction**

Designing effective digital privacy notices remains challenging. For example, too many privacy notices can lead to notice fatigue, causing a user to habitually "click through" notices rather than making informed decisions [1]. Instead, I propose that data privacy decisions can be understood as an expression of user values [2]. I also conceptually outline a system to assist users with smartphone selection based on these issues [3]. This assistant - here called a value-centered privacy assistant (VcPA) - helps create the space for users to act in accordance with their values.

In the following pages, I preliminary explore how a VcPA could enhance user value expression. To accomplish this, I utilize three user scenarios for each privacy user group - the privacy fundamentalists; the privacy pragmatists; and the privacy unconcerned - in the context of considering whether or not to download an environmental smartphone application, OpenLitterMap [4]. I then explore whether each group's value expression is preserved with the VcPA by utilizing Killmister's theory of autonomy [3,5]. To this end, I conclude that the VcPA best facilitates the value expression of privacy fundamentalists. In contrast, privacy pragmatists and the privacy unconcerned could have their value expression enhanced or hindered depending on the context and their internal states. Possible implications for future VcPA investigations are also discussed.

**Theoretical and Conceptual Background**

Designing for informed user consent in digital privacy settings is fraught with difficulties. Originally, privacy notices and policies were based around the conceptualization of users as "rational consumers" – those who weigh the service offered against their value of privacy [6]. While this view continues to inform certain policy and regulatory measures, it is now well-accepted by most privacy scholars that current notice-and-consent regimes are insufficient at providing adequate user privacy controls. For example, too many privacy notices can lead to notice fatigue, causing a user to habitually "click through" notices rather than making informed decisions [1]. In addition, "dark patterns" can coax users to consent to data collecting practices [7-9]. To combat this, privacy-preserving modifications – also called "bright patterns" – have been explored to encourage users to make better privacy choices [10,11]. These interventions, however, can be considered manipulative to the user, especially if they are unaware of a bright pattern's use [3].

Instead, I propose that we take a value-centered approach to privacy decision-making [2]. This conceptualizes data privacy as an expression of user values. When a user is faced with a privacy notice, the data collection practice of the service will either be consistent or inconsistent with a user's values. Their decision to consent or not can therefore be understood as an expression of their values.

To optimize user value expression, I propose the creation of a value-centered privacy assistant (VcPA) – an assistant that helps users select smartphone applications consistent with their values [3]. The VcPA consists of three features: suggesting alternative applications; personalized pausing; and randomized notice (summarized in Table 1). In practice, users will be prompted with personalized notices to notify them when a smartphone application's data collection practices are inconsistent with their values. While the technical details of such a personalized system have yet to be determined, the VcPA would ideally store its data locally to minimize data protection issues. Periodically, a user's values around data privacy will also be "mined" by random notices for applications previously consistent with their values. In addition, all notices will include a suggestion of alternative applications with similar functionality but more value-consistent data collection practices.

**Table 1.** Proposed features of a value-centered privacy assistant (VcPA)

| Feature | Description |
| --- | --- |
| Suggesting alternatives | On the notice itself, include suggestions for alternative applications with similar function that are consistent with the user's pre-stated values |
| Personalized pausing | Prompting a user selectively with a notice when an application is not consistent with their values |
| Randomized notices | Prompting users with notices at random time internals for applications consistent with their values |

**Methods: User Scenarios Design and Evaluation**

User scenarios are a central requirement of user-centered design. Designers can utilize scenarios as a means of translating high-level ideas into more concrete possibilities. For the purpose of this paper, I define user scenarios as "narrative descriptions" of a user's engagement with a VcPA [12]. In particular, these user scenarios have descriptive emphasis on the user's goals and values. They also reflect three privacy preference groups described elsewhere [13, 14]. These groups are: privacy fundamentalists, or users who are very concerned about disclosing their data even in the presence of privacy protections; privacy pragmatists, or users who have very specific privacy concerns about data disclosure in certain contexts; and the privacy unconcerned, or users who have mild or no concern about disclosing data, although they may still show concern for their data privacy in select circumstances [14].

In each scenario, all three hypothetical users are faced with the decision whether to download the application OpenLitterMap [4]. OpenLitterMap is a citizen science initiative that allows users to take smartphone pictures of litter and upload them into a publicly available dataset. The goal is to empower citizens to be active participants in combating local pollution. Photos of litter can be uploaded anonymously or with a username to participate in the litter "World Cup." In both cases, the system records a number of features, including time, date, location, and phone model. This means that in areas of low app use, it becomes possible to identify a user based on inference. From a value-centered privacy approach, a potential OpenLitterMap user will need to balance the value of disclosing information against the possible (albeit, small) risk of identification.

To evaluate value expression in each user scenario, I will utilize an existing systematic conception of autonomy that incorporates values. This conception of autonomy, proposed by Suzy Killmister [5], maps autonomy into four distinct dimensions: self-definition, self-realization, self-unification, and self-constitution. In the context of smartphone selection, self-definition is where a user brings together their individual goals, beliefs, and values to form a set of commitments on how to interact with smartphone applications. Self-realization consists of two states. The first internal state is when a user deliberates and decides whether to download an application based on their commitments. The second, the external state, is when a user downloads the app (or not). Self-unification is whether how the user has acted is consistent with their commitments. Self-constitution involves whether or not a user is able to modify their commitments when encountering new information about the application, such as data privacy information. From this view, then, when a user is deciding to download a smartphone application, they are involved in a dynamic process of weighing (self-realization), expressing (self-realization), and modifying (self-constitution) their defined (self-definition) values, goals, and beliefs. For it to be fully autonomous, their decision to download an application or not will also need to be consistent with their values (self-unification).

**Results and Discussion**

**User Scenario Evaluation**

**The Privacy Fundamentalist.** User #1 (the privacy fundamentalist) likes to make environmentally friendly choices. They are willing to do what they can to preserve the environment and provide the best future for their children. User #1 hears about OpenLitterMap from a friend and goes to download it. With the VcPA system, a notice appears on their screen, warning them that this application is not consistent with their personal values of security and control. They decide to check out other apps first by clicking "see alternative applications."

      At this point, there are two possible outcomes for User #1. The first is that they find a different litter clean-up application that is consistent with their values of security and control and download that one instead. In this application, the data collected may, for example, only be accessible to policy makers and environmental scientists, be encrypted, and also not collect their phone model. The second possible outcome is that User #1 does not find another application with a similar function. They may then decide to stick to their regular beach cleanings to help their environment instead of downloading an application. Without the VcPA system, User #1 may click through the privacy settings and allow the app to access their photos, camera, location, date, time, and phone model. They begin using the app when they are walking to pick up their children from school. While they want to help document litter and believe in allowing data scientists access to their documented litter data for environmental research purposes, they would feel uncomfortable if someone was able to identify their route to and from the school – and, by association, information about their children. To uphold their values of security and control, they may decide to upload their litter anonymously rather than with a username. However, they may be the only one using OpenLitterMap on that route, and it would be possible for someone looking at the data to identify them. While some may have been comfortable with this level of risk, they would not have been – they prioritize security and control over their value of environmentalism.

**The Privacy Pragmatist.** User #2 (the privacy pragmatist) has a number of practical apps on their phone. A colleague recommends that they take a look at OpenLitterMap. They go to download it.

      With the VcPA, there would be two possible outcomes for User #2. They could firstly receive a randomized notice letting them know that, while this application is consistent with their previously stated values, there is a chance of violating the values of security and control if they use the application. User #2 will then have to decide whether or not to download this application when faced with this new information. In the absence of a randomized notice, User #2 may simply click through the privacy settings and allow the app to access their photos, camera, location, date, time, and phone model, the same result without the VcPA system.

**The Privacy Unconcerned.** User #3 (the privacy unconcerned) attends a talk organized by their local Greens Club about the harmful effects of litter. The Greens Club recommends checking out OpenLitterMap. User #3 likes the idea of creating a profile to compete for the littermap "World Cup" leaderboards. They go to download the application.
Regardless of whether User #3 has the VcPA system, there is likely only one outcome. They could receive a randomized notice letting them know that, while this application is consistent with their previously stated values, there is a chance of violating the values of security and control if they use the application. They will likely then decide to download the application anyway. In the absence of the randomized notice, they will download the application. User #3 would also likely download the application without the VcPA.

**User Scenario Evaluation**

Here, I systematically assess the success of the VcPA at facilitating user value expression using the four-dimensional theory of autonomy [3,5].
In the first user scenario (privacy fundamentalist), the absence of the VcPA would have resulted in a violation of their self-unification – their actions (to download the application) would not be in alignment with their values (security and control). Thanks to personalized pausing, however, User #1 is alerted to this misalignment of their action and their values. In addition, their self-realization (acting on their beliefs) could also be enhanced if they are able to find another app using the "suggest alternatives" feature.
In the second user scenario (privacy pragmatist), User #2's autonomy may be enhanced with the VcPA. It could help with self-constitution depending on the context and their specific value preferences. When a randomized notice appears and they are presented with new information they may not have previously been aware of, they may decide to modify their values and commitments, promoting self-constitution. If, however, they do not change their values; still intend to download OpenLitterMap; and they do not download it because of the added notice, this would actually hinder self-unification because they would act in a manner inconsistent with their values. Interestingly, by introducing added friction in the form of an added notice, self-realization may also be slightly reduced by providing a small barrier to realizing their values and intention. Suzy Killmister has noted this issue previously, cautioning that interventions that encouraging a specific behavior must be consistent with what the agent has defined to uphold self-unification [5].
The same applies for the third user scenario (the privacy unconcerned). It is possible that a randomized notice could encourage them to take on new commitments concerning their privacy and thereby self-constitute; even "the privacy unconcerned" are concerned in specific circumstances [14]. Like User #2, however, it is also possible that they will suffer the same tension between self-realization/unification and self-constitution if a randomized notice changes their behavior in a manner inconsistent with their values.

**Implications for Future VcPA Design**

While the VcPA best upholds value expression of the privacy fundamentalists, the expression of the privacy pragmatists and the privacy unconcerned may be upheld depending on the context and their internal states. I have previously suggested that we could take cues from the recommender system literature to create a system of continuous exploration that minimizes the user behavioral effects of preference-mining [3,15]. The results here support that this will be critical for an effective VcPA system for the majority of users, who are privacy pragmatists [14]. The VcPA system could be optimized using user tests of privacy pragmatists to determine the right frequency and presentation of the randomized notices that maximize self-constitution while minimizing the harms to self-realization/unification.

**Concluding Thoughts**

This initial high-level conceptual exploration suggests that a VcPA could enhance or similarly preserve user value expression across different privacy groups. In order to accomplish this goal, a VcPA should be carefully designed to minimize the behavioral effects of randomized notices. Further empirical studies will also be required to further evaluate VcPA efficacy and desirability. In addition to supporting the hypothesis that a VcPA could help users make more value-centered privacy decisions, such studies will need to answer whether users will find a VcPA beneficial over current privacy controls. To validate both hypotheses, I will be utilizing a mix-method approach to elucidate relevant user values in privacy decision-making and user app download behavior with a prototype VcPA system.

**Acknowledgments.** This work is being conducted with financial support from the Science Foundation Ireland Center for Research Training in Digitally-Enhanced Reality (d-real) under Grant No. 18/CRT/6224. For the purpose of Open Access, the author has applied a CC BY public copyright license to this Author Accepted Manuscript version. Sarah is supervised by Mathieu d'Aquin (Data Science Institute, National University of Ireland Galway), Heike Felzmann (Discipline of Philosophy, National University of Ireland Galway), Kathryn Cormican (School of Engineering, National University of Ireland Galway), and Dave Lewis (ADAPT Centre, Trinity College Dublin). Sarah would also like to acknowledge Dr. Wendy Carter for reading multiple drafts of this piece and for providing invaluable feedback.

**References**

1. Schaub, F., Balebako, R, Durity, A.L., Canor, L.F.: A design space for effective privacy notices. In: Proc. 11th Symp. Usable Priv. Secur. (SOUPS), pp. 1-17. The USENIX Association, Ottawa (2015).
2. Carter, S.E.: Improving notice: the argument for a flexible, multi-value approach to privacy notice design. Presented at: 30th Intl. Assoc. Prof. Appl. Ethics (APPE) Conf. , online (2021). https://vimeo.com/509883867/c734b7c879
3. Carter, S.E.: A value-centered exploration of data privacy and personalized privacy assistants. In: CEPE/IACAP Joint Conf. Phil. Eth. AI. , Hamburg (TBP: expected June 2021).
4. OpenLitterMap, http://openlittermap.com, last accessed 2021/02/28.
5. Killmister, S.: Taking the measure of autonomy: A four-dimensional theory of self-governance. Routledge, New York (2017).
6. Hoofnagle, C.J., Urban, J.M.: Alan Westin's privacy homo economicus. Wake For. Law Rev. 14(1), 261-351 (2014).
7. Gray, C. M., Kou, Y., Battles, B., Hoggatt, J., Toombs, A.L.: The dark (patterns) side of UX design. In: Proc. Conf. Hum. Fac. Comput. Syst. (CHI), pp. 1-14. ACM, Montreal (2018).
8. Mathur, A., Acar, G., Friedman, M. J., Lucherini, E., Mayer, J., Chetty, M., Narayanan, A.: Dark patterns at scale: Findings from a crawl of 11K shopping websites. In: Proc. Conf. Comput. Sup. Coop. Work (CSCW), pp. 81:1-81:32. ACM, Austin (2019).
9. Utz, C., Degeling, M., Fahl, S., Schaub, F., Holz., T. (Un)informed consent: Studying GDPR consent notices in the field. In: Proc. Conf. Comput. Comm. Secur. (CCS), pp. 971-990. ACM, London (2019).
10. Almuhimedi, H., Schaub, F., Sadeh, N., Adjerid, I., Acquisti, A., Gluck, J., Cranor, L., Agarwal, Y. Your location has been shared 5,398 times! A field study on mobile app privacy nudging. In: Proc. Conf. Hum. Fac. Comput. Syst. (CHI), pp. 787-796. ACM, Seoul (2015).
11. Graßl, P., Schraffenberger, H., Zuiderveen Borgesius, F., Buijzen, M.: Dark and bright patterns in cookie consent requests. J. Digit. Soc. Res. 3(1), 1-38 (2021).
12. Rosson, M. B., Carroll, J.M.: Scenario-based design. In: Jacko, J., Sears. A (eds.) The Human-Computer Interaction Handbook, 1st ed., pp. 1032-1050. Lawrence Erlbaum Associates, New York (2002).
13. Westin, A.F.: Harris-Equifax consumer privacy survey. Equifax, Atlanta (1991).
14. Ackerman, M.S., Cranor, L.F., Reagle, J.: Privacy in e-commerce: Examining user scenarios and privacy preferences. In: Proc. 1st Conf. Elec. Comm., pp. 1-8. ACM, Denver (1999).

15. Jiang, R., Chiappa, S., Lattimore, T., György, A., Kohli, P. Degenerate feedback loops in recommender systems. In: Proc. Conf. AI Ethics Soc. (AIES), pp. 383-390. ACM, Honolulu (2019).